\documentclass{ws-mpla}
\usepackage[super]{cite}
\usepackage{graphicx}
\begin{document}

\markboth{Authors' names}{}

\catchline{}{}{}{}{}

\title{Modified Hawking temperature and entropy of Kerr-de Sitter black hole in Lorentz violation theory\\
}

\author{Y. Onika Laxmi }

\address{Mathematics Department, Manipur University\\
Canchipur, Manipur,  India\\}

\author{T. Ibungochouba Singh}
\address{Mathematics Department, Manipur University\\
Canchipur, Manipur,  India\\
ibungochouba@rediffmail.com}

\maketitle

\pub{Received (Day Month Year)}{Revised (Day Month Year)}

\begin{abstract}
In this paper, we discuss the tunneling of scalar particles near the event horizon of stationary and nonstationary Kerr-de Sitter black hole using Lorentz violation theory in curved space time. The modified form of Hamilton-Jacobi equation is derived from the Klein-Gordon equation by applying Lorentz violation theory. The Hawking temperatures derived from stationary and nonstationary Kerr-de Sitter black holes are modified due to Lorentz violation theory. It is noted that the change in Bekenstein-Hawking entropy and modified Hawking temperatures of stationary and nonstationary Kerr-de Sitter black hole not only depend on the black hole parameters but also on ether like vectors ${\rm u^\alpha}$.

\end{abstract}

\ccode{PACS Nos.: 04.70.Dy, 04.20.Gz, 03.65.-w.}

\section{Introduction}	

Hawkin${\rm g}^{ 1, 2}$ showed theoretically that a black hole radiates like a black body in which the temperature of radiation is dependent on the surface gravity of black hole. The discovery of Hawking radiation leads to black hole thermodynamic${\rm  s.}^{3-5}$ Since then, many scientists have proposed different techniques to study the quantum tunneling of black holes. Refs. [6,7] proposed  a new method to investigate the quantum thermal and nonthermal radiations of  stationary and nonstationary black holes by applying the tortoise coordinate transformation, Maxwell's electromagnetic field equation, Klein-Gordon equation and Dirac equation. Applying this method, many interesting results in different black holes have been derived i${\rm n.}^{8-11}$ Refs. [12-14] introduced the semiclassical tunneling technique to investigate the Hawking radiation of black hole. In this method, the Hawking radiation is taken as a tunneling process near the event horizon of the black hole and the outgoing particle produces the tunneling barrier of black hole. They obtain well-behaved coordinate system which has no singularity near the event horizon to derive the emission rate. Refs. [15-17], as an extension of Parikh and Wilczek approach, studied the Hawking radiation as a tunneling of charged massive particle at the event horizon of black hole by developing the relation between phase and group velocity of the tunneling particle. 
Applying the Hamilton-Jacobi equation, Feynman prescription and WKB approximation, Refs. [18] investigated the Hawking radiation at the event horizon of rotating and nonrotating black holes. They showed that the isotropic coordinate or invariant coordinate gives the correct Hawking temperature whereas naive coordinate leads to half of correct Hawking temperature. Ref. [19] studied the Hawking radiation at the event horizon of different black holes by using Dirac equation, Feynman prescription and Pauli Sigma matrices. In this method, the appropriate gamma matrices obtained from black hole and suitable wave function are substituted in  Dirac equations, then the action related to Boltzman factor for emission in accordance with semiclassical approximation is obtained.

Refs. [20-23] discussed the corrected Hawking temperature and entropy of black holes using first law of black hole thermodynamics and Hamilton-Jacobi equation beyond the semiclassical approximation in Schwarzschild like coordinate system and Painleve coordinate system.
Kruglo${\rm v}^{24, 25}$ proposed the quantum tunneling of boson near the event horizon of black hole using Proca equation, WKB approximation and Feynman prescription. The emission temperature of Schwarzschild background geometry which is the same as the Hawking temperature corresponding to scalar particle is also obtained. Following their work, many interesting results  have been obtained i${\rm n.}^{26-29}$

The study of Lorentz symmetry violation theory has been discussed during the past decades and many researchers have proposed different gravity models induced by Lorentz symmetry violatio${\rm n.}^{30-34}$ Refs. [35-37] proposed the Lorentz symmetry violation in flat space time using Dirac equation and ether like vectors $u^\alpha$. Since then, the Lorentz violation has been extended to curved space time by choosing ether like vectors $u^\alpha$ so that it could hold $u^\alpha u_\alpha=$ $\rm {constan{\rm t}.}^{38-42}$

The paper is outlined as follows :  In section 2, the modified Hawking temperature, heat capacity and change in entropy near the event horizon of Kerr-de Sitter black hole are derived. In section 3, the modified surface gravity and the Hawking temperature of nonstationary KdS black hole are discussed using tortoise coordinate transformation in Lorentz violation theory in curved space time. Some discussions and conclusions are given in section 4 and 5 respectively.

\section{Modified Hawking temperature of Kerr-de Sitter black hole}

The Kerr-de Sitter (KdS) solution indicates the space time geometry of a rotating black hole with non-zero cosmological constant. The vacuum solution of KdS black hole is well known Boyer-Lindquist coordinates $(t, r, \theta, \phi)$ with geometrical unit $(c=G=1)$ which is given b${\rm y}^{43}$
\begin{eqnarray}
ds^2&=&-\frac{\Delta-\Delta_\theta a^2\sin^2\theta}{R^2\Xi^2}dt^2+\frac{R^2}{\Delta_\theta}d\theta^2
-\frac{2a[\Delta_\theta(r^2+a^2)-\Delta]\sin^2\theta}{R^2\Xi^2}dtd\phi \cr&&+\frac{R^2}{\Delta}dr^2+\frac{\Delta_\theta(r^2+a^2)^2-\Delta a^2\sin^2\theta}{R^2\Xi^2}\sin^2\theta d\phi^2,
\end{eqnarray}
where
\begin{eqnarray}
&&R^2=r^2+a^2\cos^2\theta,\,\,\,\,\, \Xi=1+\frac{1}{3}\Lambda a^2,\cr
&&\Delta_\theta=1+\frac{1}{3}\Lambda a^2 \cos^2\theta,\cr
&&\Delta=(r^2+a^2)\Big(1-\frac{1}{3}\Lambda r^2 \Big)-2Mr.
\end{eqnarray}
 Eq. (1) represents rotating KdS black hole for cosmological constant $\Lambda$. The KdS black hole is well defined in the region $-\infty
\leq t\leq\infty$, $0\leq\theta\leq\pi$ and $0\leq\phi\leq2\pi$. $M$ and $a$ are the mass and rotational parameter of KdS black hole. If $\frac{1}{\Lambda}\geq M^2>a^2$, then $\Delta=0$ gives four distinct roots i.e. $ r_+,  r_h,  r_- $ and  $r_{--}$ $(r_+>r_h> r_->r_{--})$. The biggest root $r_+$ denotes the location of cosmological horizon. $r_h$ and $r_-$ represent the location of the event horizon and Cauchy horizon respectively. If $r=0$, $\theta=\frac{\pi}{2}$, then the other side of $r=0$, $r=r_{--}$ is taken as another cosmological horizo${\rm n.}^{44}$
According to Ref. [45], the event horizon of KdS black hole $r=r_h$ can be written as \begin{eqnarray}
  r_h&=&\frac{1}{\Xi}\left(1+\frac{4\Lambda M^2}{3\beta ^2\Xi}+...\right)\left(M+\sqrt{M^2-a^2 \Xi}\right), \end{eqnarray}
  where  $\beta=\sqrt{1-\frac{\Lambda}{3} a^2} $.
 There is a frame dragging effect near the event horizon of KdS black hole.
Let $\phi=\varphi-\Omega t$ and $\Omega = -\frac{g_{03}}{g_{33}}$. Then Eq. (1) reduces to
 \begin{eqnarray}
 ds^2&=&-\frac{\Delta \Delta_{\theta}R^{2}}{\Xi^{2}[\Delta_{\theta}(r^2+a^2)^2-\Delta a^2 \sin^2\theta]}dt^2+\frac{R^2}{\Delta}dr^2+\frac{R^2}{\Delta_{\theta}}d\theta^2\cr&&+\frac{[\Delta_{\theta}(r^2+a^2)^2-\Delta  a^2 \sin^2\theta]\sin^2\theta}{R^{2}\Xi^{2}}d\phi^2, 
\end{eqnarray}
where the angular velocity at the event horizon of KdS black hole is given as
\begin{eqnarray}
 \Omega &=&\frac{a}{r_h^2+a^2}.
 \end{eqnarray}
 According to Refs. [46, 47], the surface gravity of KdS black hole at the event horizon $r=r_h$ is derived as 
  \begin{eqnarray}
  \kappa=\lim_{ g_{00} \rightarrow 0}\Bigg(-\frac{1}{2}\sqrt{-\frac{g^{11}}{g_{00}}}\frac{d g_{00}}{dr}\Bigg)
=\frac{(r_h- M -\frac{2}{3}\Lambda r_h^3-\frac{1}{3}\Lambda a^2 r_h)}{\Xi(r_h^2+a^2)}.
  \end{eqnarray}
  The Hawking temperature of KdS black hole is connected with surface gravity via $T_H=\frac{\kappa}{2\pi}$ as
  \begin{eqnarray}
  T_H=\frac{1}{2 \pi}\Bigg[\frac{(r_h- M -\frac{2}{3}\Lambda r_h^3-\frac{1}{3}\Lambda a^2 r_h)}{\Xi(r_h^2+a^2)}\Bigg].
  \end{eqnarray}
  To discuss heat capacity near the event horizon of black hole, the mass parameter KdS black hole might be obtained from 
$\Delta(r_h)=0$ as

\begin{eqnarray}
M=\frac{r_h}{2}+\frac{a^2}{2r_h}-\frac{\Lambda r_h^3}{6}-\frac{\Lambda a^2 r_h}{6}.
\end{eqnarray}
The heat capacity $(C_h)$ of black hole is defined by
\begin{eqnarray}
C_h&=&\frac{\partial M}{\partial T_H}=\Bigg(\frac{\partial M}{\partial r_h}\Bigg)\Bigg(\frac{\partial r_h}{\partial T_H}\Bigg).
\end{eqnarray}
The heat capacity near the event horizon of KdS black hole is calculated as
  \begin{eqnarray}
C_{h} =\dfrac{\partial M}{\partial T_H}&=& \frac{2\pi \Xi (r_{h}^{2}+a^{2})^{2}[3r_{h}^{2}-3a^{2}-\Lambda r_{h}^{2}(3r_{h}^{2}+a^{2})]}{3(a^4-r_{h}^{4})+4a^2r_{h}^{2}(3-2\Lambda r_{h}^{2})-\Lambda r_{h}^{2}(a^4 +3r_{h}^{4})}.
\end{eqnarray}
The modified form of Hamilton-Jacobi equation in Lorentz theory in curved space time is given b${\rm y}^{48}$ 
\begin{eqnarray}
(g^{\mu\nu}+\lambda u^\mu u^\nu)\partial_\mu I\partial_\nu I+m^2 &=&0.
\end{eqnarray}
where $ \lambda $ and $u^\alpha$ are the correction parameter and ether like vectors respectively.
As $ \lambda $ tends to zero in the above equation, the Lorentz violation theory is cancelled and the original Hamilton-Jacobi equation in curved space time is obtained.
The ether like vectors $u^\alpha$ are constant in the flat space time of the canonical coordinate system. The ether like vectors $u^\alpha$ are not constant in curved space time. But we can take the vectors $u^\alpha$ from curved space time that satisfies the condition $u^\alpha u_\alpha= {\rm constant}$.  To investigate the change in entropy of stationary KdS black hole, we can choose $u^\alpha$ from (4) that satisfies $u^\alpha u_\alpha= {\rm constant}$ and $u^\alpha$ are related to coordinate system acquired by the space time. The expressions of $ u^t, u^r, u^\theta $ and $u^\phi$ are defined by
  \begin{eqnarray}
   u^t&=&\frac{c_t}{\sqrt{-g_{tt}}} = \frac{c_t \Xi \sqrt{\Delta_{\theta}(r^2+a^2)^2-\Delta a^2 \sin^2\theta}}{\sqrt{\Delta \Delta_{\theta}R^{2}}} ,\cr
u^{r}&=&\frac{c_{r}}{\sqrt{g_{rr}}} =c_r\sqrt{\frac{\Delta}{R^2}},\cr
u^{\theta}&=&\frac{c_{\theta}}{\sqrt{g_{\theta\theta}}} =c_\theta \sqrt{\frac{\Delta_\theta}{R^2}},\cr
 u^{\phi}&=&\frac{c_{\phi}}{\sqrt{g_{\phi\phi}}} =\frac{c_\phi \Xi \sqrt{R^2}}{\sqrt{[\Delta_{\theta}(r^2+a^2)^2-\Delta a^2 \sin^2\theta}]\sin^2\theta},\end{eqnarray}
 where $c_t$, $c_r$, $c_{\theta}$ and $c_{\phi}$ are arbitrary constants. $u^\alpha$ satisfies the condition
    
\begin{eqnarray}
u^{\alpha}u_{\alpha}=-c_t^2+c_r^2+c_\theta^2+c_\phi^2={\rm constant}.
\end{eqnarray}
 
   Using Eqs. (12) and (4) in Eq. (11), the dynamical equation of scalar particle with mass $m$ in stationary KdS black hole is obtained as
    \begin{eqnarray}
 && g^{00}(\frac{\partial I}{\partial t})^2+g^{11}(\frac{\partial I}{\partial r})^2+g^{22}(\frac{\partial I}{\partial \theta})^2+g^{33}(\frac{\partial I}{\partial \phi})^2+\lambda u^\mu u^\nu\partial_\mu I\partial_\nu I+m^2=0.
   \end{eqnarray}
   It is known that the above equation involves the variables $t, r ,\theta$ and $\phi$. To separate the variables on the Hamilton principal functions $I$, we can choose the action $I$ as follows
  \begin{eqnarray} 
 I&=&-\omega t +S(r,\theta) +j \phi +\delta,
  \end{eqnarray}
  where $S(r, \theta)$, $\omega$ and $j$ are the generalized momentum, particle energy and angular momentum along the $\phi$-axis respectively and $\delta$ is a complex constant.
 Using Eq. (15) in Eq. (14), a quadratic equation in $\frac{\partial S}{\partial r}$ is obtained as
  \begin{eqnarray}
  A\Bigg(\frac{\partial S}{\partial r}\Bigg)^2+B \left(\frac{\partial S}{\partial r}\right)+C=0.
  \end{eqnarray}
  Then the two roots of the above equation are given by
  \begin{eqnarray}
   S= \int \frac{-B\pm\sqrt{B^2-4AC}}{2A} dr, 
  \end{eqnarray}
 where \begin{eqnarray}
 A&=&g^{11}+\lambda u^r u^r,\cr
 B&=& 2\lambda u^r u^\theta \left(\frac{\partial I}{\partial\theta}\right)-2\lambda u^t u^r (\omega -j\Omega) +2\lambda u^r u^\phi j,\cr
 C&=&(g^{00}+\lambda u^t u^t)(\omega -j\Omega)^2 +(g^{22}+\lambda u^\theta u^\theta)\left(\frac{\partial I}{\partial \theta}\right)^2+(g^{33}+\lambda u^\phi u^\phi)j^2\cr&&-2\lambda u^t(\omega -j\Omega) \Big(u^\theta\left(\frac{\partial I}{\partial\theta}\right)+ u^\phi j\Big)+2\lambda u^\theta u^\phi j \left(\frac{\partial I}{\partial\theta}\right)+m^2.
 \end{eqnarray}
 Applying residue theorem of complex analysis and Feynman prescription near the event horizon of KdS black hole, the integration of Eq. (17) can be written as
  \begin{eqnarray}
 S_\pm=\frac{i\pi  \Xi(r_h^2+a^2)\left[\lambda c_t c_r \pm \sqrt{1-\lambda c_t^2 +\lambda c_r^2}\right](w-j\Omega)}{2(1+\lambda c_r^2)\{(1-\frac{\Lambda}{3} a^2) r_h-M-\frac{2\Lambda r_h^3}{3}\}},
  \end{eqnarray}
  where $S_+$ and $S_-$ are the outgoing particle and ingoing particle respectively.
  The probabilities which cross the black hole near the event horizon are given by
  \begin{eqnarray}
  \Gamma_{emission} &=&{\rm exp} (-2{\rm Im I})= {\rm exp}[-2({\rm Im S_+} +{\rm Im }\delta)] 
  \end{eqnarray}
  and \begin{eqnarray}
  \Gamma_{absorption} &=&{\rm exp} (-2{\rm Im I})= {\rm exp}[-2({\rm Im S_- }+{\rm Im} \delta)].
  \end{eqnarray}
  There is a $100\%$ chance the ingoing particle to enter the KdS black hole according to WKB approximation. This indicates that ${\rm Im S_+} = -{ \rm Im S}_-$.
  We calculate the probability of outgoing particle as 
  \begin{eqnarray}
  \Gamma _{rate} =\frac{\Gamma _{emission}}{\Gamma_{absorption}} &=&{\rm exp} \Bigg[-\frac{ 2\gamma\Xi \pi (r_h^2+a^2) (\omega-j \Omega)}{\{(1-\frac{\Lambda}{3} a^2) r_h-M-\frac{2\Lambda r_h^3}{3}\}} \Bigg],
  \end{eqnarray}
  where $\gamma=\frac{\sqrt{1-\lambda c_t^2 +\lambda c_r^2}}{1+\lambda c_r^2}$. 
  Eq. (22) is similar to Boltzmann factor according to semiclassical approximation. The Hawking temperature near the event horizon of KdS black hole in Lorentz violation theory is given by 
  \begin{eqnarray}
  T=\frac{\{(1-\frac{\Lambda}{3} a^2) r_h-M-\frac{2\Lambda r_h^3}{3}\}}{2  \gamma \pi \Xi(r_h^2+a^2)}.
  \end{eqnarray}

  If $\lambda=0$ in the above equation, the Lorentz violation has been cancelled. In such case Eq. (23) is consistent with Eq. (7). If $\gamma > 1$ or $\gamma <1$, the Hawking temperature decreases or increases due to the presence of correction term $\lambda$ and ether like vectors $u^\alpha$.
  
  The modified heat capacity at the event horizon of KdS black hole is obtained as
  
\begin{eqnarray}
C^{'}_{h} &=&\dfrac{\partial M}{\partial T}
= \frac{2\pi\gamma \Xi (r_{h}^{2}+a^{2})^{2}[3r_{h}^{2}-3a^{2}-\Lambda r_{h}^{2}(3r_{h}^{2}+a^{2})]}{3(a^4-r_{h}^{4})+4a^2r_{h}^{2}(3-2\Lambda r_{h}^{2})-\Lambda r_{h}^{2}(a^4 +3r_{h}^{4})}.
\end{eqnarray}
 As $\gamma$ tends to unity, Eq. (24) is consistent with original heat capacity given in Eq. (10). If $\gamma>1$ or 
 $\gamma<1$, the heat capacity increases or decreases near the event horizon of KdS black hole depending upon the choices of correction term $\lambda$ and ether like vectors $u^\alpha$.
  
  Using Eqs. (3) and (5) in Eq. (19) for the outgoing particle, we get
  \begin{eqnarray}
  {\rm Im S}&=& \frac{\gamma^{'}}{2}\left[\frac{\frac{\pi  k_1^2 k_2^2}{\Xi }}{\frac{\beta^2}{\Xi}k_1 k_2 -M-\delta}\omega+\frac{\pi \Xi a^2}{\frac{\beta^2}{\Xi}k_1 k_2 -M-\delta}\omega-\frac{\pi \Xi a}{\frac{\beta^2}{\Xi}k_1 k_2 -M-\delta}j\right], 
  \end{eqnarray}
  where 
  \begin{eqnarray}
   \gamma^{'}&=&\frac{\lambda c_t c_r +\sqrt{1-\lambda c_t^2 +\lambda c_r^2}}{1+\lambda c^2_r },\cr
  k_1&=&\left(1+\frac{4\Lambda M^2}{3\beta ^2\Xi}+...\right),\cr
  k_2&=&\left(M+\sqrt{M^2-a^2 \Xi}\right),\cr
  \delta&=&\frac{2\Lambda}{3 \Xi ^3}\left(1+\frac{4\Lambda M^2}{3\beta ^2\Xi}+...\right)^3\left(M+\sqrt{M^2-a^2 \Xi}\right)^3.
  \end{eqnarray}
  
  To obtain the biggest value of the integration, we ignore second order terms of KdS black hole mass parameter at the numerator and denominator in Eq. (25). Then we find as
   \begin{eqnarray}
 {\rm Im S}&=&  \frac{\gamma^{'}}{2}\left[\frac{\pi   k_2^2}{\beta^2\left[k_2 -\frac{M\Xi}{\beta^2}\right]}\omega+\frac{\pi \Xi^2 a^2}{\beta^2\left[ k_2 -\frac{M\Xi}{\beta^2}\right]}\omega-\frac{\pi \Xi^2 a}{\beta^2\left[k_2-\frac{M\Xi}{\beta^2}\right]}j\right].
   \end{eqnarray}
   Taking the self-gravitational interaction into account, the mass parameter KdS black hole is allowed to fluctuate. If a black hole emits a particle $\omega$ and angular momentum $j$, the KdS black hole parameter will be $M-\omega$ and $J-j$ respectively.
   To find the change in  Bekenstein-Hawking entropy of KdS black hole, the term $(1-\frac{\Xi}{\beta^2}){\rm M}$ is ignored. Then we get
  \begin{eqnarray}
  {\rm Im S}= \frac{\gamma^{'}}{2}\left[\int^\omega_0\frac{\pi   k_2^2}{\beta^2\sqrt{M^2-a^2\Xi} }d\omega^{'}+\int^\omega_0\frac{\pi \Xi^2 a^2}{\beta^2\sqrt{M^2-a^2\Xi}}d\omega^{'}
  -\int^j_0\frac{\pi \Xi^2 a}{\beta^2\sqrt{M^2-a^2\Xi}}dj^{'}\right].\cr\
   \end{eqnarray} 
   Changing  $ M $ by $M-\omega$ and $j$ by $J-j$ and putting the value of $k_2$ in the above equation, we obtain
   \begin{eqnarray}
  {\rm Im S}&=& \frac{\gamma^{'}}{2}\Bigg[\frac{-\pi }{\beta^2}\int^{M-\omega}_M\frac{  k_2^2}{\sqrt{(M'-\omega')^2-a^2\Xi} }d(M'-\omega^{'})\cr&&-\frac{\pi \Xi^2 a^2}{\beta^2}\int^{M-\omega}_M\frac{1}{\sqrt{(M'-\omega')^2-a^2\Xi}}d(M'-\omega^{'})\cr&&
  +\frac{\pi \Xi^2 a}{\beta^2}\int^{J-j}_j\frac{1}{\sqrt{(M'-\omega')^2-a^2\Xi}}d(J-j')\Bigg],
   \end{eqnarray} 
   where $J-j=(M-\omega)a$.   
  The imaginary part of the action finally yields
   \begin{eqnarray} 
  {\rm Im S}&=&\frac{-\gamma^{'}\pi }{2\beta^2 }\Bigg[\int^{M-\omega}_M\frac{2(M'-\omega')^2+2(M'-\omega')\sqrt{(M'-\omega')^2-a^2 \Xi}}{\sqrt{(M'-\omega')^2-a^2 \Xi}}d(M'-\omega^{'})\cr&&-\int^{M-\omega}_M \frac{a^2 \Xi}{\sqrt{(M'-\omega')^2-a^2 \Xi}}d(M'-\omega^{'})\Bigg].
   \end{eqnarray}
   Calculating the $\omega^{'}$ integral, that gives
   \begin{eqnarray}
   {\rm Im S}&=&\frac{-\gamma^{'}\pi }{2\beta^2 }\Bigg[(M-\omega)\sqrt{(M-\omega)^2-a^2\Xi}+(M-\omega)^2-M\sqrt{M^2-a^2\Xi}-M^2\Bigg],\cr
   &=&\frac{-\gamma^{'}\pi }{4\beta^2}\Bigg[\Bigg((M-\omega)+\sqrt{(M-\omega)^2-a^2\Xi}\Bigg)^2-\Bigg( M+\sqrt{M^2-a^2\Xi}\Bigg)^2\Bigg],\cr
   &=& \frac{-\gamma^{'} \pi}{2}(r_f^2-r_i^2).
   \end{eqnarray}
   Using WKB approximation, the tunneling rate is obtained as
   \begin{eqnarray}
   \Gamma &\sim &exp(-2 \rm Im S),\cr
   &= & exp [\gamma^{'} \pi (r_f^2-r_i^2)],\cr
   &= & exp [\gamma^{'} \Delta S_{BH}],
    \end{eqnarray}

   where $\gamma' \Delta S_{BH} =\gamma' (S_{BH}(M-\omega)-S_{BH}(M))$ is the modified entropy of KdS black hole in Lorentz violation theory. $ r_{i}=\frac{1}{\sqrt{2}\beta }\Bigg[M+\sqrt{M^2-a^2\Xi}\Bigg]$ and  $ r_{f}=\frac{1}{\sqrt{2}\beta }\Bigg[(M-\omega)+\sqrt{(M-\omega)^2-a^2\Xi}\Bigg]$
 are the locations of horizons before and after emission of particle.
 If $\lambda=0$, the Lorentz violation theory has been cancelled, the original change in Bekenstein-Hawking entropy is obtained. When $\gamma^{'}>1$, the change in Bekenstein-Hawking entropy increases and $\gamma^{'}<1$, the change in Bekenstein-Hawking entropy decreases near the event horizon of KdS black hole. In above cases, the change in Bekenstein-Hawking entropy depends on correction parameter $\lambda$ and ether like vectors $u^\alpha$. 
\section{Nonstationary rotating KdS black hole}

The metric of rotating nonstationary KdS black hole in retarded time coordinate${\rm s}^{10}$$(u, r, \theta, \phi)$ is defined by 
 \begin{eqnarray}
 ds^2&=&\frac{1}{R^2 \Xi^2}[\Delta_\lambda-\Delta_\theta a^2\sin^2\theta]du^2-\frac{R^2}{\Delta_\theta}d\theta^2+\frac{2a}{R^2 \Xi^2}[\Delta_\theta(r^2+a^2)-\Delta_\lambda]\sin^2\theta du d\phi\cr&&+\frac{2}{\Xi}[du-a\sin^2\theta d\phi]dr-\frac{1}{R^2 \Xi^2} [\Delta_\theta(r^2+a^2)^2-\Delta_\lambda a^2 \sin^2\theta]\sin^2\theta d\phi^2,
 \end{eqnarray}
 where $R^2, \Xi, \Delta_\theta $ are given in Eq. (2). The term $\Delta_\lambda $ is given by
 \begin{eqnarray}
&&\Delta_\lambda=r^2+a^2-2M (u)r-\frac{1}{3}\Lambda r^2(r^2+a^2),
  \end{eqnarray}
where $M(u)$ is the mass of nonstationary KdS black hole. The location of event horizon of stationary and nonstationary KdS black hole can be obtained from null surface equation $F(u, r, \theta, \phi)=0.$
  The expression of null surface equation is given by  \begin{eqnarray}
  g^{\mu \nu}\frac{\partial F}{\partial x^\mu}\frac{\partial F}{\partial x^\nu}=0.
\end{eqnarray}
The location of event horizon of stationary and nonstationary black holes can be obtained from the null surface equation (35) using generalized tortoise coordinate transformation. The space time geometry outside the event horizon of nonstationary black hole can be described by the tortoise coordinate and in such case, $r_*$ tends to positive infinity when tending to infinite point and $r_*$ tends to negative infinity at the event horizon of black hole. To study the Hawking radiation of nonstationary black hole, the tortoise coordinate transformation is defined b${\rm y}^{49-54}$
  
\begin{eqnarray}
    r_{*}&=&r+\frac{1}{2\kappa(u_{0},\theta_{0},\phi_{0})}ln{\frac{r-r_{h}(u,\theta,\phi)}{r_{h}(u,\theta,\phi)}},\cr
    u_{*} &=&u-u_{0} ,\,\,\theta_{*}=\theta-\theta_{0} ,\,\,\phi_{*} =\phi-\phi_{0},
\end{eqnarray}
   where $u_0, \theta_0 $ and $\phi_0$ are the arbitrary constants under the coordinate transformation. From the above equation, we get
   
\begin{eqnarray}
   \frac{\partial}{\partial r}&=&\Big[{1+\frac{1}{2\kappa(r-r_h) }}\Big] \frac{\partial}{\partial r_{*}},\cr
  \frac{\partial}{\partial u}&=&\frac{\partial}{\partial u_{*}}-\frac{r r_{h,u}}{2\kappa r_{h}(r-r_h)}\frac{\partial}{\partial r_{*}},\cr
   \frac{\partial}{\partial \theta}&=&\frac{\partial}{\partial \theta_{*}}-\frac{r r_{h,\theta}}{2\kappa r_{h}(r-r_h)}\frac{\partial}{\partial r_{*}},\cr
   \frac{\partial}{\partial\phi}&=&\frac{\partial}{\partial \phi_{*}}-\frac{r r_{h,\phi}}{2\kappa r_{j}(r-r_h)}\frac{\partial}{\partial r_{*}},
 \end{eqnarray}
where $r_{h,u}=\frac{\partial r_h}{\partial u}$,  $r_{h,\theta}=\frac{\partial r_h}{\partial \theta}$ and  $r_{h,\phi}=\frac{\partial r_h}{\partial \phi}$. $r_{h,u}$ represents the evaporation rate at the event horizon of KdS space time. If $\frac{\partial r_h}{\partial u}>0$, the event horizon of KdS space time is expanded (absorbing black hole ) and when $\frac{\partial r_h}{\partial u}<0$, the event horizon of KdS space time is contracted. $\kappa\equiv\kappa(u_0, \theta_0, \phi_0)$ is taken as surface gravity of KdS space time which depends on retarded time and angular coordinates. Using Eqs. (33) and (36) in Eq. (35) and taking $r\rightarrow r_h$, the horizon equation of nonstationary KdS black hole is derived as
    \begin{eqnarray}
    &&\frac{a^2\Xi^2 \sin^2\theta r^2_{h,u}}{\Delta_\theta}+2(r_h^2+a^2)\Xi r_{h,u}+\frac{2a \Xi^2 r_{h,u}r_{h,\phi}}{\Delta_\theta}+2a\Xi r_{h,u} \cr&&+\Delta_\lambda(r_h) +\Delta_\theta r^2_{h,\theta} +\frac{\Xi^2 r^2_{h,\phi}}{\Delta_\theta \sin^2\theta}=0,
       \end{eqnarray}
where $\Delta_\lambda (r_h)=r^2_h +a^2-2M(u)r_h-\frac{1}{3}\Lambda r^2_h(r^2_h +a^2)$. It is noted that the location of event horizon of nonstationary black hole varies with retarded time $u=t-r_*$ and angular co-ordinates $\theta, \phi$.
       From Eqs. (11) and (33), the dynamical equation of scalar particle with mass $m$ in curved space time is obtained as
   \begin{eqnarray}
 && g^{00}\Big(\frac{\partial I}{\partial u}\Big)^2+2g^{01}\Big(\frac{\partial I}{\partial u}\Big)\Big(\frac{\partial I}{\partial r}\Big)+2g^{03}\Big(\frac{\partial S}{\partial u}\Big)\Big(\frac{\partial I}{\partial \phi}\Big)+g^{11}\Big(\frac{\partial I}{\partial r}\Big)^2+2g^{13}\Big(\frac{\partial I}{\partial r}\Big)\Big(\frac{\partial I}{\partial \phi}\Big)\cr&&+g^{22}\Big(\frac{\partial I}{\partial \theta}\Big)^2+g^{33}\Big(\frac{\partial I}{\partial \phi}\Big)^2+\lambda u^\mu u^\nu\partial_\mu I\partial_\nu I+ m^2=0.
   \end{eqnarray}
   Since ether like vectors are not constant in curved space time, we can choose ${\rm u^\alpha}$ from nonstationary space time Eq. (33) so that we can make $u^\alpha u_\alpha={\rm constant}$. The ether like vectors ${\rm u^\alpha}$ are related to the properties of black hole and system of coordinate adopted by the metric space. The ether like vectors ${\rm u^\alpha}$ are choosen as  
   \begin{eqnarray}
   u^u&=&\frac{k_u}{\sqrt{g_{uu}}}=\frac{k_u \sqrt{R^2}\Xi}{\sqrt{\Delta_\lambda-\Delta_\theta a^2\sin^2\theta}},\cr
u^r&=&\frac{k_r\sqrt{g_{uu}}}{g_{ur}} =\frac{k_r\sqrt{\Delta_\lambda-\Delta_\theta a^2\sin^2\theta}}{\sqrt{R^2}},\cr
u^{\theta}&=&\frac{k_{\theta}}{\sqrt{g_{\theta\theta}}} =k_\theta \sqrt{\frac{-\Delta_{\theta}}{R^2}},\cr
u^\phi&=&\frac{\Delta_{\lambda}k_{\phi}}{\sqrt{g_{\phi\phi}}} =\frac{\Delta_{\lambda}k_\phi \sqrt{R^2}\Xi} {\sqrt{-[\Delta_\theta(r^2+a^2)^2-\Delta_\lambda a^2 \sin^2\theta]\sin^2\theta}},
    \end{eqnarray}
  where $k_u$,  $k_r$, $k_\theta$ and $k_\phi$ are arbitrary constants. Using Eqs. (37) and (40) in Eq. (39), we get
  \begin{eqnarray}
    \frac{D}{E}\Big(\frac{\partial I}{\partial r_*}\Big)^2+2\Big(\frac{\partial I}{\partial u_*}\Big)\Big(\frac{\partial I}{\partial r_*}\Big)+2\frac{F}{E}\Big(\frac{\partial I}{\partial r_*}\Big)+ 2\kappa (r-r_h)\frac{G}{E}=0,
    \end{eqnarray}
    where 
    \begin{eqnarray}
     D&=&\frac{1}{2\kappa (r-r_h)r^2_h}\Big[(g^{00}+\lambda u^u u^u)r^2r^2_{h,u}-2r_h\Big\{(g^{01}+\lambda u^u u^r)rr_{h,u}+(g^{13}\cr&&+\lambda u^\phi u^r) rr_{h,\phi}+\lambda u^ru^\theta rr_{h,\theta}\Big\}\{ 2k(r-r_h)+1\}+2(g^{03}+\lambda u^u u^\phi)r^2r_{h,u}r_{h,\phi}\cr&&+r^2_h(g^{11}+\lambda u^ru^r)\lbrace2\kappa(r-r_h)+1\rbrace^2+(g^{22}+\lambda u^\theta u^\theta )r^2r^2_{h,\theta}\cr&& +(g^{33}+\lambda u^\phi u^\phi)r^2r^2_{h,\phi}+2\lambda u^\theta r^2r_{h,\theta}(u^u r_{h,u}+u^\phi r_{h,\phi})\Big],\cr
    E&=&-(g^{00}+\lambda u^u u^u)\frac{rr_{h,u}}{r_h}+(g^{01}+\lambda u^u u^r)\lbrace2k(r-r_h)+1\rbrace\cr&&-(g^{03}+\lambda u^u u^\phi)\frac{rr_{h,\phi}}{r_{h}} -\frac{\lambda u^u u^\theta rr_{h,\theta}}{r_h},\cr
   F&=&-(g^{03}+\lambda u^u u^{\phi})\frac{rr_{h,u}p_{\phi}}{r_{h}}+\{g^{13}p_{\phi}+\lambda u^{r}( u^{\phi}p_{\phi}+u^{\theta}p_{\theta})\}\lbrace2k(r-r_h)+1\rbrace \cr&&-(g^{22}+\lambda u^{\theta}u^{\theta})\frac{rr_{h,\theta}p_{\theta}}{r_{h}}-(g^{33}+\lambda u^{\phi}u^{\phi})\frac{rr_{h,\phi}p_{\phi}}{r_{h}}-\frac{\lambda u^{u} u^{\theta}rr_{h,u}p_{\theta}} {r_{h}}\cr&&-\frac{\lambda u^{\theta} u^{\phi} rr_{h,\phi}p_{\theta}}{r_h}-\frac{\lambda u^{\theta} u^{\phi} rr_{h,\theta}p_{\phi}}{r_h},\cr
   G&=& g^{00}\omega^2-2g^{03}p_\phi \omega +g^{22}p^2_\theta +g^{33}p^2_\phi +\lambda u^u u^u \omega^2 -2\lambda u^u u^\phi \omega p_\phi +\lambda u^\theta u^\theta p^2_\theta\cr&& +\lambda u^\phi u^\phi p^2_\phi -2\lambda u^u u^\theta \omega p_\theta +2\lambda u^\theta u^\phi p_\theta p_\phi +m^2. 
    \end{eqnarray}
    To study the Hawking temperature, the action ${\rm I}$ in Eq. (15) can be writen as  
 $ {\rm I}=-\omega u_{*}+I_{0}(r_{*},\theta_{*},\phi_{*})$, then  we get
  
    \begin{eqnarray}    
    \frac{\partial I}{\partial u_*}=-\omega,\,\,\,\frac{\partial I}{\partial \theta_*}= p_{\theta},\,\,\,\frac{\partial I}{\partial \phi_*}= p_{\phi},
   \end{eqnarray}
where $\omega$ is the energy of emitted scalar particle. $p_\theta$ and $p_\phi$ are the components of generalized momenta of scalar particle along the angular coordinates $\theta$ and $\phi$ respectively. To obtain surface gravity and Hawking temperature at the event horizon of nonstationary  KdS black hole, we assume that the coefficient of $\Big(\frac{\partial I}{\partial r_*}\Big)^2$ approaches to unity as $r\rightarrow r_h$, $u\rightarrow u_0$, $\theta\rightarrow\theta_0$ and $\phi\rightarrow \phi_0$. From Eq. (41), an infinite limit of the form $\frac{0}{0}$ is obtained near the event horizon of KdS black hole. Using L'Hopital's rule, the surface gravity is obtained as
\begin{eqnarray} 
\kappa = \frac{r_h (1+2\Xi r_{h,u})-M-\frac{2}{3}\Lambda r^3_{h}-\frac{\Lambda}{3}a^2 r_h - r^{-1}_{h}\lbrace\Delta_{\lambda} +\Xi (r_h^2+a^2)r_{h,u}+a\Xi r_{h,\phi}\rbrace}{\lbrace\Xi (r_h^2+a^2)+\frac{a^2\Xi^2\sin^2\theta r_{h,u}}{\Delta_{\theta}}\rbrace(1+2r_{h,u})+Z+\lambda Y}, \cr\   
  \end{eqnarray} 
  
where $Z$ and $Y$ are given by
\begin{eqnarray} 
Z&=&2\Delta_{\theta}r^2_{h,\theta}+r_{h,\phi}\Big(\frac{4a\Xi^2 r_{h,u}}{\Delta_{\theta}}+\frac{a\Xi^2}{\Delta_{\theta}}+\frac{2\Xi^2 r_{h,\phi}}{\Delta_{\theta}\sin^2\theta}+2a\Xi\Big), \cr
   Y&=&\frac{k^2_u \rho^{4}r_{h,u}\Xi^2}{\Delta_{\theta}a^2\sin^2\theta}+\rho^2 k_u k_r \Xi-\frac{\rho^2 k_u k_{\theta}\Xi r_{h,\theta}}{a\sin\theta}.
     \end{eqnarray}  
Using Eq. (43) in Eq. (41) and taking $r\rightarrow r_h$, then we obtain
     \begin{eqnarray}
     \Bigg(\frac{\partial I}{\partial r_*}\Bigg)^2+2(\omega-\omega_0) \frac{\partial I}{\partial r_*}=0.
     \end{eqnarray}    
     The value of chemical potential, $\omega_0$ near the event horizon of nonstationary KdS black hole  is 
\begin{eqnarray}
w_0&=&\frac{1}{g^{01}-g^{00}r_{h,u}-g^{03}r_{h,\phi}-\lambda u^u u^u r_{h,u}+\lambda u^r u^r-\lambda u^u u^\phi r_{h,\phi}-\lambda u^u u^\theta r_{h,\theta}}\cr && \times[g^{13}p_\phi-g^{03}r_{h,u}p_\phi -g^{22}r_{h,\theta}p_\theta -g^{33}r_{h,\phi}p_\phi-\lambda u^u r_{h,u}(u^\phi p_\phi+u^\theta p_\theta)\cr&& - \lambda(u^\theta r_{h,\theta}+ u^\phi r_{h,\phi})( u^\theta p_\theta + u^\phi p_\phi) +\lambda  u^r (u^\theta p_\theta+u^\phi p_\phi)].
     \end{eqnarray}
     Eq. (47) represents the chemical potential of nonstationary KdS space time due to tortoise coordinate transformation (36). The chemical potential, $ w_0$ depends on the black hole mass, cosmological constant, generalized momenta of scalar particle, retarded time, angular coordinates, correction parameter $\lambda$ and ether like vectors $u^\alpha$. If $\lambda$ and $u^\alpha$ tend to zero, Eq. (47) is consistent with earlier literatures [10, 46]. From the solution of Eq. (37), we obtain
     \begin{eqnarray}
     \frac{\partial I}{\partial r_*}=\frac{[2\kappa (r-r_h) +1](\omega-\omega_0)\pm (\omega-\omega_0)}{[2\kappa (r-r_h)]}.
     \end{eqnarray}
     It is observed that Eq. (48) has a singularity near the event horizon of KdS black hole. Integrating Eq. (48) by applying residue theorem of complex analysis and Feynman prescription, the imaginary part of the radial action $I$ is derived as
     \begin{eqnarray}
    {\rm Im I\pm}=\frac{\pi}{2\kappa}[(\omega-\omega_0)\pm (\omega-\omega_0)],
     \end{eqnarray}
     where ${\rm I_+}$ and ${\rm I_-}$ represent the outgoing scalar particle and ingoing scalar particle at the event horizon of KdS black hole. Taking outgoing and ingoing of scalar particle, the tunneling probability which crosses at the event horizon of KdS black hole is calculated as
     \begin{eqnarray}
     \Gamma=\frac{\Gamma_{emission}}{\Gamma_{absorption}}={\rm exp} [-\frac{(\omega-\omega_0)}{T}].
     \end{eqnarray}
      The modified Hawking temperature near the event horizon of nonstationary KdS black hole due to Lorentz violation theory is given by 
 \begin{eqnarray}
 T&=&T_h(1+\lambda H)^{-1}\cr&=&T_h(1-\lambda H+\lambda^{2}H^{2}-\lambda^3 H^3+...),
\end{eqnarray}
where the value of $T_h$ and $H$ are  
\begin{eqnarray}
T_h &=&\frac{1}{2\pi}\Big[\frac{r_h (1+2\Xi r_{h,u})-M-\frac{2}{3}\Lambda r^3_{h}-\frac{\Lambda}{3}a^2 r_h }{\lbrace\Xi (r_h^2+a^2)+\frac{a^2\Xi^2\sin^2\theta r_{h,u}}{\Delta_{\theta}}\rbrace(1+2r_{h,u})+Z}\cr&&-\frac{r^{-1}_{h}\lbrace\Delta_{\lambda} +\Xi (r_h^2+a^2)r_{h,u}+a\Xi r_{h,\phi}\rbrace}{\lbrace\Xi (r_h^2+a^2)+\frac{a^2\Xi^2\sin^2\theta r_{h,u}}{\Delta_{\theta}}\rbrace(1+2r_{h,u})+Z}\Big]
  \end{eqnarray}
   and 
   \begin{eqnarray}
   H=\frac{\frac{k^2_u \rho^{4}r_{h,u}\Xi^2}{\Delta_{\theta}a^2\sin^2\theta}+\rho^2 k_u k_r \Xi-\frac{\rho^2 k_u k_{\theta}\Xi r_{h,\theta}}{a\sin\theta}}{\lbrace\Xi (r_h^2+a^2)+\frac{a^2\Xi^2\sin^2\theta r_{h,u}}{\Delta_{\theta}}\rbrace(1+2r_{h,u})+Z}.
   \end{eqnarray}
   
 From Eqs. (52) and (53), it is known that  the Hawking temperature near the event horizon of KdS black hole is modified due to Lorentz violation theory. The modified Hawking temperature $(T)$ of nonstationary KdS black hole depends not only on the mass of the black hole but also on the properties of event horizon, cosmological constant $\Lambda$, retarded time $u$, correction term $\lambda$ and on the ether like vectors $u^\alpha$.
If $ k_u \rho^2 r_{h_{,u}}\Xi+\Delta_{\theta}a\sin\theta(a\sin\theta  k_r-k_{\theta}r_{h,\theta})=0$, then $H\longrightarrow 0 $. In such case, our result is consistent with the earlier literature${\rm s.}^{10, 11, 46}$

\section{Discussion}
First the line element of stationary KdS black hole is transformed into static form using frame dragging effect given in Eq. (4).
\begin{figure} [h]
\centering
\centering{\includegraphics[width=350pt] {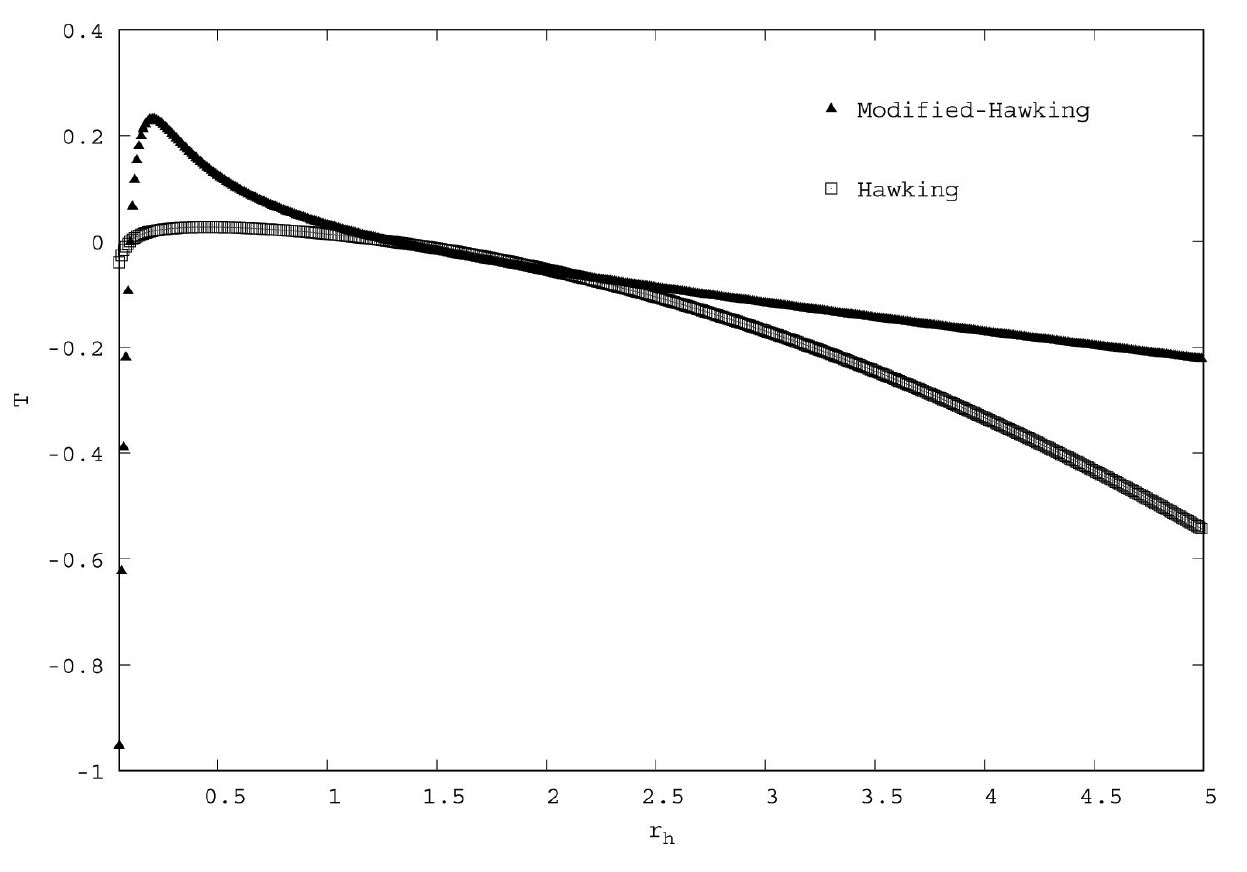}}
\caption{Plot of original and modified Hawking temperature with radius of event horizon, $r_{h}$ of KdS black hole. Here $a=0.1, \Lambda=0.6, \lambda=0.005, c_r=1.5, c_t=0.4$.}
\end{figure} 
Using modified form of Hamilton-Jacobi equation, Feynman prescription and WKB approximation, the modified Hawking temperature, heat capacity and change in Bekenstein-Hawking entropy are derived in Eq. (23), Eq. (24) and Eq. (31) respectively. It is noted that both are dependent on correction term $\lambda$ and ether like vectors $u^\alpha$.

If $\gamma >1$, the modified Hawking temperature near the event horizon of KdS black hole decreases and if $\gamma <1$,  the modified Hawking temperature increases in Eq. (23). If $\gamma=1$, the modified heat capacity (24) approaches to original heat capacity (10).

If $\gamma>1$, the modified heat capacity increases and if  $\gamma <1$, the modified heat capacity is smaller than that of original heat capacity near the event horizon of stationary KdS black hole.
   
 The change in Bekenstein-Hawking entropy near the event horizon of KdS black hole increases or decreases  if $\gamma'>1$ or $\gamma'<1$. When $\lambda=0$ and $c_t=c_r$, $\lambda\neq0,$  the Lorentz violation has been cancelled and the original change in Bekenstein-Hawking entropy near the event horizon of KdS black hole is recovered.

     The Hawking temperature of rotating nonstationary KdS black hole is also investigated using Klein-Gordon equation, generalized tortoise coordinate transformation and L'Hopital rule in Lorentz violation theory. According to Damour and Ruffin${\rm i}^{6}$ and Sanna${\rm n}^{7}$, the traditional coordinate transformation is given by
\begin{eqnarray}
r_* &=& r+\frac{1}{2\kappa_1}{\rm ln}(r-r_h(u, \theta, \phi)),\cr  
     u_{*} &=&u-u_{0} ,\,\,\theta_{*}=\theta-\theta_{0} ,\,\,\phi_{*} =\phi-\phi_{0}.
\end{eqnarray}
 Using Eq. (54) in Eq. (41), the modified surface gravity and Hawking temperature of nonstationary rotating KdS black hole are \begin{eqnarray}
 \kappa_{1}=\frac{r_h (1+2\Xi r_{h,u})-M-\frac{2}{3}\Lambda r^3_{h}-\frac{\Lambda}{3}a^2 r_h}{{\lbrace\Xi (r_h^2+a^2)+\frac{a^2\Xi^2\sin^2\theta r_{h,u}}{\Delta_{\theta}}\rbrace(1+2r_{h,u})+Z+\lambda Y} }
\end{eqnarray} and
 \begin{eqnarray}
      T_1=\frac{1}{2\pi}\frac{r_h (1+2\Xi r_{h,u})-M-\frac{2}{3}\Lambda r^3_{h}-\frac{\Lambda}{3}a^2 r_h}{{\lbrace\Xi (r_h^2+a^2)+\frac{a^2\Xi^2\sin^2\theta r_{h,u}}{\Delta_{\theta}}\rbrace(1+2r_{h,u})+Z+\lambda Y} }.
 \end{eqnarray}
      In Eqs. (51) and (56) , if we put $\lambda$ = $p_\theta$ = $p_\phi$ = $0$ and $ k_u \rho^2 r_{h_{,u}}\Xi+\Delta_{\theta}a\sin\theta(a\sin\theta  k_r-k_{\theta}r_{h,\theta})=0$, Eqs. (51) and (56) are concordant with the literatur${\rm e}.^{46}$ Eq. (54) gives another chemical potential of nonstationary rotating KdS black hole in Lorentz violation theory as   
\begin{eqnarray}
w_p&=&\frac{1}{g^{01}-g^{00}r_{h,u}-g^{03}r_{h,\phi}-\lambda u^u u^u r_{h,u}+\lambda u^r u^r-\lambda u^u u^\phi r_{h,\phi}-\lambda u^u u^\theta r_{h,\theta}}\cr && \times[g^{13}p_\phi-g^{03}r_{h,u}p_\phi -g^{22}r_{h,\theta}p_\theta -g^{33}r_{h,\phi}p_\phi-\lambda u^u r_{h,u}(u^\phi p_\phi+u^\theta p_\theta)\cr&& - \lambda(u^\theta r_{h,\theta}+ u^\phi r_{h,\phi})( u^\theta p_\theta + u^\phi p_\phi) +\lambda  u^r (u^\theta p_\theta+u^\phi p_\phi)].
     \end{eqnarray}
It is observed that Eq. (57) is consistent with Eq. (47).
     The surface gravities derived from Eq. (44) and (55) can be combined as
     \begin{eqnarray}
      \kappa &= &\kappa_1 +\Xi_1,
      \end{eqnarray}
       where $\kappa$ and $\kappa_1$ represent the surface gravities of nonstationary KdS black hole due to tortoise coordinate transformations (36) and (54).
     $\Xi_1$ indicates the constant term due to tortoise coordinate transformation (36) and its expression is 
     \begin{eqnarray}
     \Xi_1&=&-\frac{r^{-1}_{h}\lbrace\Delta_{\lambda} +\Xi (r_h^2+a^2)r_{h,u}+a\Xi r_{h,\phi}\rbrace}{\lbrace\Xi (r_h^2+a^2)+\frac{a^2\Xi^2\sin^2\theta r_{h,u}}{\Delta_{\theta}}\rbrace(1+2r_{h,u})+Z+\lambda Y}.
     \end{eqnarray}
     The rate of correction for Hawking temperature near the event horizon of nonstationary KdS black hole is given by
     \begin{eqnarray}
     \delta&=& -\frac{r^{-1}_{h}\lbrace\Delta_{\lambda} +\Xi (r_h^2+a^2)r_{h,u}+a\Xi r_{h,\phi}\rbrace}{r_h (1+2\Xi r_{h,u})-M-\frac{2}{3}\Lambda r^3_{h}-\frac{\Lambda}{3}a^2 r_h }.
     \end{eqnarray}
It is worth mentioning that the correction rate is independent of correction term $\lambda $ and the ether like vectors $u^\alpha$ but depends on black hole mass $M$, rotational parameter $a$, angular coordinate $\theta$, cosmological constant $\Lambda$ and generalized momenta. For stationary KdS black hole in the absence of Lorentz violation theory, Eq. (51) and (56) reduce to the same Hawking temperature as
    \begin{eqnarray}
    T_H= T_1=\frac{1}{2\pi}\Bigg[\frac{r_h-M-\frac{2}{3}\Lambda r^3_{h}-\frac{\Lambda}{3}a^2 r_h}{\Xi (r_h^2+a^2)}\Bigg].
    \end{eqnarray}
    It is noted that for the stationary KdS space time, the different tortoise coordinate transformations give the same Hawking temperature in the absence of Lorentz violation theory which is exactly equal to the actual calculation given in Eq. (7). From Eq. (58), $\kappa_1$ approaches to zero, the Hawking temperature $T$ does not tend to zero due to extra term $\Xi_1$. If $\Xi_1$ approaches to zero in Eq. (58), Eq. (44) is consistent with Eq. (55). From Eqs. (47) and (57), we observe that the chemical potential derived from different tortoise coordinate transformations are equal near the event horizon of black hole. It can be concluded that the tortoise coordinate transformation given in Eq. (36) is more suitable and accurate in the study of modified surface gravity near the event horizon of nonstationary KdS space time in Lorentz violation theory. From Eqs. (44) and (55), the different surface gravities are obtained using the different tortoise coordinate transformations near the event horizon of nonstationary KdS black hole in Lorentz violation theory.

\section{Conclusion}

 In this paper, the tunneling of scalar particle near the event horizon of stationary KdS black hole is investigated using Klein-Gordon equation in Lorentz violation theory, Feynman prescription and WKB approximation. Then the corresponding Hawking temperature, heat capacity and change in Bekenstein-Hawking entropy near the event horizon are derived. The Hawking temperature, heat capacity and change in entropy are modified due to presence of correction term $\lambda$ and ether like vectors $u^\alpha$. 
      
    The modified surface gravities of nonstationary rotating black hole are also studied using different tortoise coordinate transformations in Lorentz violation theory. Using null surface equation and tortoise coordinate transformation, the horizon equation of nonstationary KdS black hole is obtained. The modified surface gravities and the modified Hawking temperatures are derived with the help of event horizon equation. It is known that the modified Hawking temperature depends not only on the correction term $\lambda$ but also on the ether like vectors $u^\alpha$.
    If we use Eq. (36) in the study of surface gravity and Hawking radiation of black hole, a constant term $\Xi_1$ is seen to be appeared in the expressions of surface gravity and Hawking temperature near the event horizon of KdS black hole. If $\Xi_1$ tends to zero, the two modified surface gravities and Hawking temperatures are equal. If $\lambda$ and $\Xi_1$ approach to zero, the original surface gravities near the event horizon of nonstationary black hole are recovered and are concordant with the earlier literature${\rm s.}^{8, 9, 54}$ It is also seen that the correction rate $\delta$ of Hawking temperature in Lorentz violation theory does not depend on correction term $\lambda$ and the ether like vectors $u^\alpha$. The different tortoise coordinate transformations yield the same chemical potential at the event horizon of black hole but the values of surface gravities and Hawking temperatures are different. This shows that Eq. (36) is more reliable and accurate in the study of modified surface gravity near the event horizon of nonstationary of KdS black hole. For the stationary space time,  the different methods yield the same Hawking temperature and the change in Bekenstein-Hawking entropy in the absence of Lorentz violation theory.
\section*{Acknowledgments}
The first author acknowledges the DST INSPIRE, New Delhi, India for providing financial support (Grant No. IF190759).

\end{document}